\documentclass[10pt,twocolumn]{article}
\usepackage[margin=0.75in]{geometry}
\usepackage{booktabs,graphicx,url,hyperref,amsmath,microtype}
\usepackage[numbers,sort&compress]{natbib}
\usepackage{times}
\hypersetup{colorlinks,linkcolor=blue,citecolor=blue,urlcolor=blue}
\title{When the Model Retires: An Empirical Study of LLM Migration\\ in Open-Source Applications}
\author{Hyungjin Lukas Kim\thanks{Corresponding author: \texttt{kimhj@mju.ac.kr}.}\\
\normalsize Department of Future and Convergence Business Administration, Myongji University, Seoul, Republic of Korea}
\date{}
\begin{document}
\maketitle

\begin{abstract}
Applications built on commercial large language model (LLM) APIs depend on model versions that providers retire on their own schedule, with notice periods ranging from one year to two weeks. We ask what actually happens to applications when a model is retired. We mine GitHub for commits that migrate away from officially deprecated models and endpoints of OpenAI, Anthropic, and Google, matching each commit to the provider's published announcement and shutdown dates. From 22,555 commits in 17,703 non-fork repositories (2024--2026), 5,139 are matched to an official event; two independent coders validated a stratified sample of 300 ($\kappa$ = 0.89--0.95), and we reweight all estimates by their labels. We find that an estimated 82\% (95\% CI 79--84) of migrations away from retired models were committed \emph{after} the shutdown date---after the application had started failing---regardless of repository popularity, prior retirement experience, or the presence of a provider-abstraction layer. The share tracks the provider's notice policy: 89\% for Anthropic's 60--114-day notices versus 13\% for OpenAI's one-year Assistants API notice, and each $e$-fold increase in notice length reduces the odds of post-shutdown migration by about three quarters. Model identifiers are hard-coded in 94\% of migrating applications, migration effort scales from a median of 6 added lines for prompt-only applications to nearly 700 for fine-tuned ones, and only 8\% of migrations switch provider. We release the dataset and pipeline and discuss implications for deprecation policy, dependency-risk assessment of LLM products, and tooling.
\end{abstract}

\section{Introduction}
Software that calls a commercial LLM API depends on a service with an expiry date that the provider controls. Unlike a library, a retired model cannot be pinned and kept: on the shutdown date every request fails. Trackers report that model lifecycles have shortened from 18--24 months to 6--12 months, and 2026 saw the largest deprecation notice to date, when OpenAI scheduled GPT-3.5 Turbo, GPT-4, GPT-4 Turbo, o1, o3-mini, and o4-mini for a single-day shutdown on 23 October 2026, and shut down the Assistants API on 26 August 2026 after a one-year notice. Anthropic retired every Claude~3.x model between October 2025 and April 2026 on 60--114 day notices. Google shut down the Gemini~2.0 Flash family on 1 June 2026.

Prior work has documented the \emph{difficulty} of migrating between models---prompt re-engineering in a single case study~\cite{Tursio2025}, regression tooling~\cite{RETAIN2024}, an end-of-life migration framework~\cite{EOL2026}, and evidence that aggregate scores hide item-level regressions~\cite{ItemLevel2026}---but not what happens across the ecosystem when a retirement actually lands: how many applications migrate before the deadline, how many break, and what the migration costs in code. Studies of API deprecation in library ecosystems~\cite{Robbes2012,Sawant2018,Decan2018,Kula2018} established that clients frequently ignore deprecation warnings; LLM retirement is a harsher form of the same phenomenon, because ignoring it is not an option.

We contribute (1) a dataset of migration commits matched to an official event table compiled from the three providers' deprecation pages; (2) the first ecosystem-scale measurement of reactive (post-shutdown) versus proactive migration; (3) a comparison across providers' notice policies and a dose--response estimate; (4) a keyword-and-manual taxonomy of breakage; and (5) evidence on migration size and on architectural adaptation (abstraction layers, fallbacks, evaluation suites).

\section{Background: Deprecation Policies}
\label{sec:policy}
\textbf{OpenAI} distinguishes deprecation (announcement) from shutdown and promises at least six months' notice for generally available models, three months for specialized variants, and ``as little as two weeks'' for preview models. \textbf{Anthropic} promises at least 60 days' notice for publicly released models, publishes ``not sooner than'' retirement floors for active models, and has committed to preserving retired weights. \textbf{Google} publishes ``earliest'' shutdown dates for GA Gemini models and at least two weeks for previews; announcement dates are not published, so for Google we measure timing only relative to shutdown. Table~\ref{tab:events} summarizes the events used in this study (full table with sources in the replication package).

\begin{table}[t]\centering\footnotesize\setlength{\tabcolsep}{3pt}
\caption{Selected official retirement events (announce $\to$ shutdown).}
\label{tab:events}
\begin{tabular}{@{}llrl@{}}\toprule
Provider & Model / endpoint & Notice (d) & Shutdown \\\midrule
OpenAI & Assistants API & 365 & 2026-08-26 \\
OpenAI & gpt-4-vision-preview & 183 & 2024-12-06 \\
OpenAI & gpt-4o-realtime-preview & 234 & 2026-05-07 \\
OpenAI & gpt-4-turbo-preview & 181 & 2026-03-26 \\
OpenAI & o1-mini & 182 & 2025-10-27 \\
OpenAI & o1-preview & 91 & 2025-07-28 \\
OpenAI & GPT-3.5/4/4-Turbo, o1, o3/o4-mini & 184 & 2026-10-23$^{*}$ \\
Anthropic & claude-3-5-sonnet & 76 & 2025-10-28 \\
Anthropic & claude-3-7-sonnet & 114 & 2026-02-19 \\
Anthropic & claude-3-5-haiku & 62 & 2026-02-19 \\
Anthropic & claude-3-haiku & 60 & 2026-04-20 \\
Anthropic & claude-sonnet-4 / opus-4 & 62 & 2026-06-15 \\
Google & gemini-2.0-flash(-lite) & n/a & 2026-06-01 \\
Google & gemini-3-pro-preview & n/a & 2026-03-09 \\\bottomrule
\end{tabular}
\smallskip\\\parbox{\linewidth}{\scriptsize $^{*}$Pending at the time of writing; excluded from timing analyses.}
\end{table}

\section{Method}
\textbf{Corpus.} We queried the GitHub commit-search API with 22 query strings combining retired model identifiers or endpoint names with migration vocabulary (\emph{deprecated}, \emph{retired}, \emph{migrate}, \emph{shutdown}, \emph{model\_not\_found}), over quarterly windows (2024--mid 2025) and monthly windows (mid 2025--September 2026), up to 200--300 results per window. Forks were excluded and commits de-duplicated by SHA, yielding 22,555 unique commits in 17,703 repositories (TypeScript 28\%, Python 27\%, JavaScript 6\%). Repository metadata (stars, forks, archived flag, primary language) and per-commit diff statistics (added/deleted lines, changed files) were retrieved through the GraphQL API for 22,513 commits, and changed file paths with patch hunks through the REST API for all 3,933 commits tied to retired events (3,931 retrievable).

\textbf{Event matching.} We compiled an event table from the providers' official deprecation pages summarized in Section~\ref{sec:policy} (30 OpenAI, 9 Anthropic, 7 Google entries). Google publishes an \emph{earliest possible} shutdown date for GA models at release---one year after release for the Gemini~2.0 family---and later fixes the exact date; because the date of that later announcement is not archived on the page, we exclude Google from the notice-length regression in the main analysis and report a sensitivity analysis that assigns GA models a 365-day soft notice and previews 14 days. A commit is matched to the most specific model or endpoint pattern that appears in its message with word boundaries (short identifiers such as \texttt{o1} require delimiters on both sides). To retain only commits that move \emph{away} from the model, we require a migration verb (\emph{migrate, replace, switch, upgrade, update, remove, drop, swap, retire, deprecate, sunset, shut down}) and exclude commits more than 30 days before the announcement, AI-tool authorship headers that embed a model identifier, and dependency-bot commits. We further exclude commits in which the retired identifier appears as the \emph{destination} of an arrow or in which the named destination is an older model (e.g., ``gemini-1.5-flash $\to$ gemini-2.0-flash''), since these migrate \emph{into} the model; this removed 33\% of candidates, concentrated in Google events. This yields 5,139 matched commits (4,544 repositories); 3,933 concern events already shut down and are used for timing analyses. Two coders (graduate researchers in the author's group, working independently and blind to each other's labels and to the automated flags) labelled a stratified random sample of 300 matched commits (50 per provider $\times$ pre/post shutdown) on six codes: genuine migration away from the retired model, nature (production repair, planned change, catalogue cleanup), breakage, silent failure, pinned snapshot as cause, and abstraction layer (introduced, pre-existing, none). Agreement was high on five codes (Cohen's $\kappa$ = 0.95, 0.89, 0.94, 0.94, 0.92) and moderate on pinned snapshots ($\kappa$ = 0.62), which we therefore report only as a lower bound. Only 62\% of matched commits were genuine migrations by both coders' judgement; the rest were catalogue or documentation refreshes, migrations \emph{into} the model that survived the direction filter, or unrelated commits that mention the identifier. Crucially, the genuine rate differs by stratum---24\% to 80\% before shutdown versus 48\% to 86\% after, and lowest for Google---so we reweight all population shares by the stratum-specific genuine rate and report bootstrap confidence intervals. Against the agreed human labels, the breakage keyword rule has precision 0.81 and recall 0.92; the silent-failure rule 0.57/0.57; the abstraction rule 0.70/0.42; and the pinned-snapshot rule 0.25/0.83, so we do not report keyword-based pinned-snapshot rates.

\textbf{Timing.} For each matched commit we compute days from announcement and from shutdown. A migration is \emph{reactive} if committed after the shutdown date. Because recently retired events are right-censored, dose--response analyses use only events retired at least 28 days before collection and commits within $[-365,+180]$ days of shutdown.

\textbf{Message coding.} Keyword classifiers flag breakage (\emph{404, not found, fail, broke, error, crash}), silent failure (\emph{failing silently, returned null/empty, caught the exception}), pinned snapshots, abstraction layers (\emph{LiteLLM, OpenRouter, LangChain, router, gateway, fallback, provider-agnostic}), evaluation (\emph{eval, regression, test suite}), prompt and parameter changes (\emph{max\_completion\_tokens, temperature}), and cost.

\section{Results}
\subsection{RQ1: When do applications migrate?}
\begin{figure}[t]\centering\includegraphics[width=\linewidth]{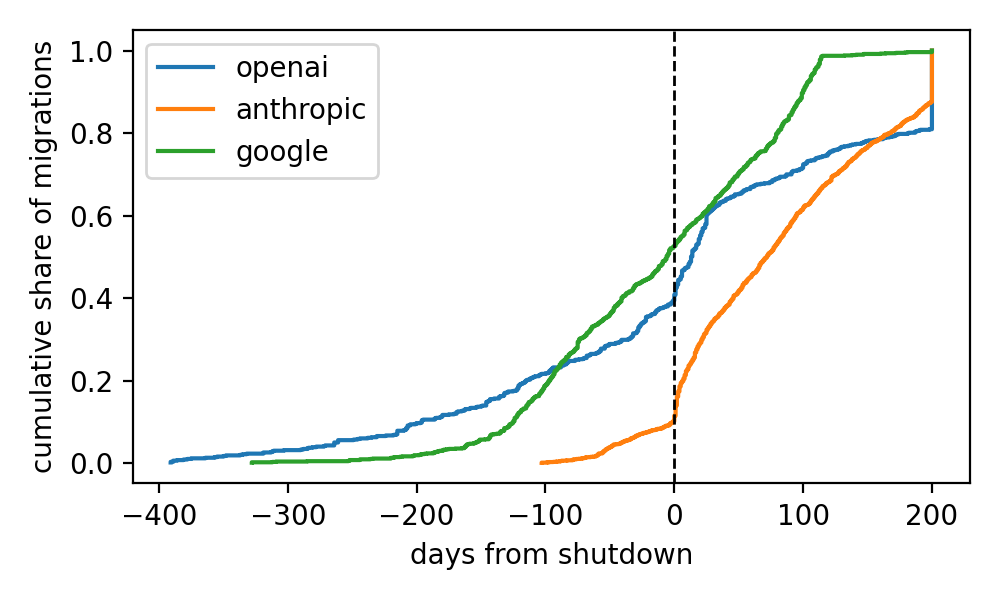}
\caption{Cumulative share of migration commits by days from official shutdown (retired events). The dashed line is the shutdown date; the share to its left is proactive migration.}\label{fig:cdf}\end{figure}

Across the 3,933 commits tied to already-retired events, 72\% were committed after the shutdown date; after reweighting by the human-validated genuine rate, which is lower before shutdown than after, the estimate is \textbf{82\% (95\% CI 79--84)}. The median migration landed 39 days after shutdown and 158 days after the announcement. A calendar-based ``after shutdown'' is not the same as ``after breaking'': among genuine post-shutdown migrations the coders labelled 40\% as production repairs, 45\% as planned changes with no breakage mentioned, and 14\% as catalogue cleanup, reflecting soft shutdowns, applications not in active use, and messages that omit the symptom. Figure~\ref{fig:cdf} shows the cumulative distribution by provider. Repository popularity has no detectable association with timing: the raw post-shutdown share is 72\% for zero-star repositories, 71\% for 1--9 stars, 71\% for 10--99, and 74\% for 100+ stars. Because many popular repositories are model catalogues, gateways, and frameworks that prune a dead identifier rather than repair a broken feature, we re-computed the tiers after excluding catalogue-like projects (repository names matching \emph{models, catalog, registry, router, gateway, proxy, sdk, client, provider}, or migrations touching five or more identifier-bearing files; 16\% of commits): the reweighted shares are 82\%, 80\%, 81\%, and 82\% across the four tiers. Star count is therefore not a proxy for retirement readiness. Reactive fixes also arrive slowly: only 9\% of them land within a week of shutdown, 27\% within a month, and 55\% within three months, so a retired model typically leaves an application broken for weeks. For short-notice retirements the arrival is a wave: the week after the Claude~3.5 Haiku, 3 Haiku, and Opus~4.1 shutdowns each carried 25 times the weekly volume of fix commits seen in the eight weeks before, against 2.1$\times$ for the Assistants API and 1.3$\times$ for Gemini~2.0 Flash.

\subsection{RQ2: Does the notice policy matter?}
\begin{figure}[t]\centering\includegraphics[width=\linewidth]{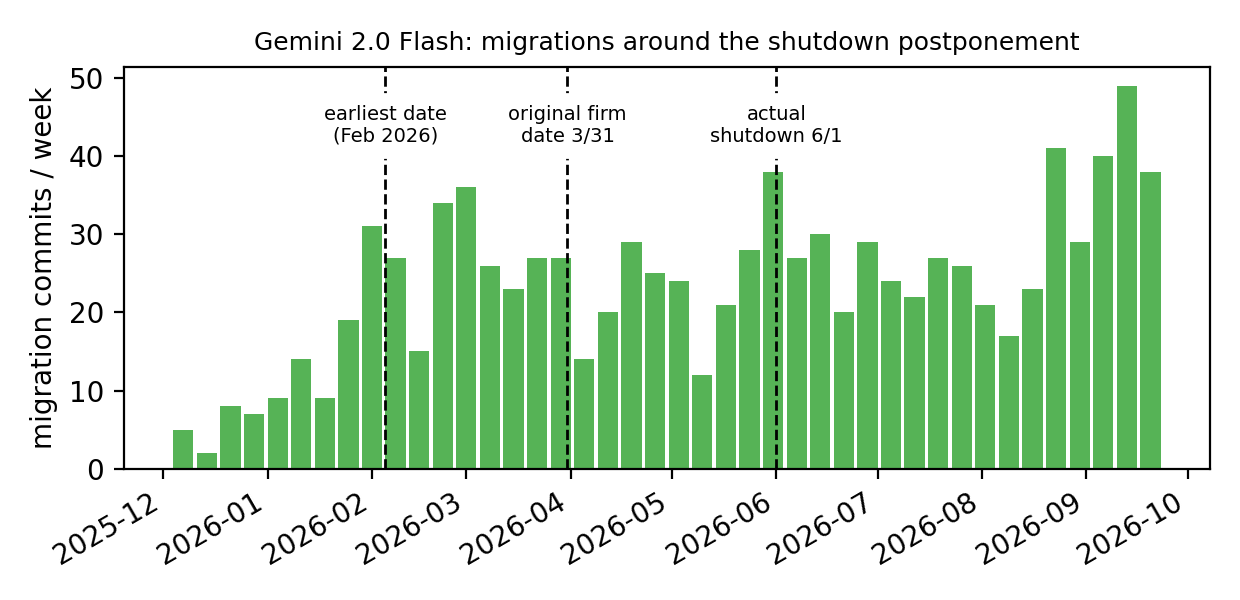}
\caption{Weekly migration commits away from Gemini~2.0 Flash(-Lite). Dashed lines mark the ``earliest'' date published at release, the original firm shutdown date, and the actual shutdown.}\label{fig:gemini}\end{figure}

\begin{figure}[t]\centering\includegraphics[width=\linewidth]{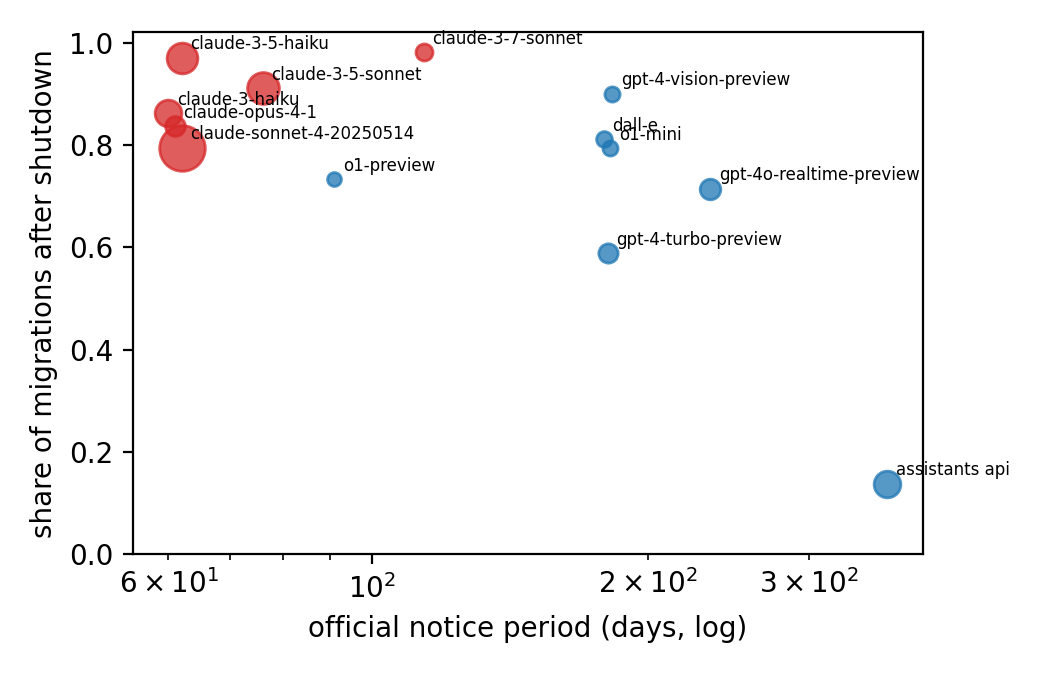}
\caption{Reactive share versus official notice period, per event (marker size $\propto n$).}\label{fig:dose}\end{figure}

Table~\ref{tab:prov} contrasts providers. Reweighted by the human-validated genuine rate, 89\% (CI 88--91) of Anthropic-related migrations were reactive, against 66\% (CI 59--73) for OpenAI and 64\% (CI 52--78) for Google. The raw shares (86/50/47\%) understate OpenAI and especially Google, whose pre-shutdown candidates were mostly false positives (Google's Gemini~2.0 Flash absorbed many migrations \emph{into} it from Gemini~1.5 that survived the direction filter). Google's Gemini~2.0 Flash listed an ``earliest'' shutdown from release and the date was later moved to 1~June~2026; its genuine migrations still straddle the shutdown date, consistent with a long but soft notice. At event level (Table~\ref{tab:ev}), the 60--76-day Anthropic retirements of Claude~3 Haiku, 3.5 Haiku, and 3.5 Sonnet saw 86--98\% reactive migration, whereas the one-year Assistants API notice saw 13\%. Figure~\ref{fig:dose} plots reactive share against notice length per event; the event-level Spearman correlation is $\rho=-0.53$ ($p=0.06$, $n=13$ events). A commit-level logistic regression of reactive migration on $\log(\text{notice days})$ with repository-clustered standard errors gives a coefficient of $-1.52$ (SE 0.09, $p<0.001$) on the raw matched set and $-1.36$ (SE 0.09, $p<0.001$) when each commit is weighted by its stratum's human-validated genuine rate: each $e$-fold increase in notice reduces the odds of post-shutdown migration by roughly 74--78\%. Adding Google under the soft-notice assumption described in Section~3 yields $-0.95$ (SE 0.06, $p<0.001$), weaker but in the same direction, consistent with a soft one-year date being less effective than a firm one. (The 300-commit validation sample itself is stratified on the outcome and therefore cannot be used to re-estimate this regression directly.) Google's Gemini~2.0 Flash offers a small natural experiment: its firm shutdown date was moved from 31~March to 1~June~2026 in early March. Figure~\ref{fig:gemini} shows no discontinuity at either date---migrations trickled in at a near-constant 23--29 per week from February through July, with only a modest dip (28.5 to 23.5 per week) in the four weeks after the postponement became known and no spike in the week after the actual shutdown, in contrast to the 25-fold spikes after Anthropic's firm 60-day retirements. A soft, movable date thus appears to produce neither a deadline effect before shutdown nor a breakage wave after it; whether that reflects genuinely gradual shutdown behaviour on Google's side or lower dependence on the model among our sampled repositories, we cannot tell from commits alone. Two caveats: the Assistants API is an endpoint whose migration requires rewriting rather than renaming, which may itself encourage planning; and Claude~Sonnet~4 (62-day notice, 79\% reactive) shows that a short notice on a \emph{current-generation} model draws somewhat faster reaction than on an older one, but still leaves four in five applications migrating after the fact.

\begin{table}[t]\centering\footnotesize\setlength{\tabcolsep}{3.5pt}
\caption{Reactive migration by provider (events already retired; robust window $[-365,+180]$ d, retired $\ge$28 d before collection).}
\label{tab:prov}
\begin{tabular}{lrrrr}\toprule
Provider & $n$ & Reactive raw / reweighted & Med.\ days & Breakage \\\midrule
Anthropic & 1,763 & 86\% / 89\% & +72 & 54\% \\
Google & 1,104 & 47\% / 64\% & $-6$ & 28\% \\
OpenAI & 553 & 50\% / 66\% & +14 & 29\% \\\bottomrule
\end{tabular}
\end{table}

\begin{table}[t]\centering\footnotesize\setlength{\tabcolsep}{3.5pt}
\caption{Reactive share by event ($n\ge 20$, robust window).}
\label{tab:ev}
\begin{tabular}{llrrr}\toprule
Event & Notice (d) & $n$ & Reactive & Breakage \\\midrule
claude-3-haiku & 60 & 218 & 86\% & 55\% \\
claude-opus-4-1 & 61 & 92 & 84\% & 50\% \\
claude-3-5-haiku & 62 & 308 & 97\% & 58\% \\
claude-sonnet-4 & 62 & 750 & 79\% & 59\% \\
claude-3-5-sonnet & 76 & 335 & 91\% & 43\% \\
claude-3-7-sonnet & 114 & 53 & 98\% & 57\% \\
dall-e-2/3 & 179 & 37 & 81\% & 22\% \\
gpt-4-turbo-preview & 181 & 85 & 59\% & 44\% \\
o1-mini & 182 & 29 & 79\% & 17\% \\
gpt-4-vision-preview & 183 & 113 & 97\% & 35\% \\
gpt-4o-realtime-preview & 234 & 105 & 71\% & 43\% \\
Assistants API & 365 & 217 & 14\% & 21\% \\
gemini-2.0-flash(-lite) & $\ge$365$^\dagger$ & 1,053 & 47\% & 28\% \\\bottomrule
\end{tabular}
\smallskip\\\parbox{\linewidth}{\scriptsize $^\dagger$Shutdown date published at release; announcement date not published; excluded from regression.}
\end{table}

\subsection{RQ3: Breakage}
By keyword, 41\% of matched commits describe breakage (48\% post-shutdown, 29\% pre-shutdown). Among human-validated genuine migrations the coders found breakage in 63\% of post-shutdown Anthropic migrations, 29\% of Google's, and 19\% of OpenAI's, and in 42\%, 33\%, and 4\% respectively before shutdown; the provider ordering mirrors the notice ordering. Manual reading of the sample surfaced four recurring forms. \emph{Hard failure}: the endpoint returns 404/\texttt{not\_found\_error} and the feature stops (``every outreach draft 404'd after discovery succeeded''). \emph{Silent failure} (8\% of genuine migrations by human coding, 14 of 180; the keyword rule finds the same prevalence with precision 0.57): exception handlers converted the retired-model error into an apologetic chat message, a null result, a demo-mode fallback, or---in one case---a constant answer (``breed identification always returned `Golden Retriever'''). One commit reports that ``AI analysis had been failing silently in production'' for weeks after the Gemini~2.0 Flash shutdown. \emph{Parameter incompatibility}: the replacement model rejects \texttt{max\_tokens} or non-default \texttt{temperature}, so a rename alone does not restore service (7.5\% of commits mention parameter changes). \emph{Pinned snapshots}: dated identifiers such as \texttt{claude-3-5-sonnet-20241022}, pinned for reproducibility, were named as the proximate cause in a small number of commits (6 of 118 genuine migrations on which both coders agreed); a keyword rule that flags any dated identifier overstates this pattern by an order of magnitude, so we treat it as present but uncommon.

\subsection{RQ4: Effort and where the model lives}
Migrations are small: median 2 changed files (75th percentile 5), 14 added and 6 deleted lines. File-level data for 3,931 retired-event migrations sharpen this picture (Table~\ref{tab:files}). Only \textbf{6\% of migrations touch configuration or documentation files alone}; 94\% edit source, i.e., the model identifier was hard-coded. These file-level shares are robust to the precision problem noted in Section~3: reweighting by stratum genuine rate gives 6.4\%, and restricting to the 181 human-validated genuine migrations with file data gives 1.7\%, so hard-coding is if anything more prevalent among true migrations. The identifier had to be changed in a median of one file, but in \textbf{26\% of migrations in three or more files}, and 16\% of migrations also touch a test file (12\% for prompt-only applications).

Using patch content as a proxy for architecture, 78\% of migrating applications are prompt-only, 13\% show tool or agent calls, 8\% retrieval (embeddings, vector stores), and 2\% fine-tuning. Effort scales steeply with architecture: prompt-only migrations change a median of 2 files and 6 lines; agent applications 5 files and 63 lines; RAG 7 files and 155 lines; fine-tuned applications 14 files and 693 lines, because the replacement base model invalidates the tuned checkpoint. The ordering and magnitudes hold on the human-validated subset (prompt-only 2 files and 7 lines; agent, RAG, and fine-tuned 8--9 files and 224--411 lines). Reactive shares are similar across architectures (58--74\%), so heavier applications are not more proactive---they merely pay more when the retirement lands.

Abstraction layers are rare and mostly retrofitted. Only \textbf{3\% of migrations occur in code that already routed calls through an abstraction} (LiteLLM, OpenRouter, a gateway, a model map; 3\% reweighted, 7\% among human-validated migrations); those migrations are the smallest non-trivial ones (4 files, 20 lines) but are no less reactive (70\%). A further \textbf{4\% introduce such a layer} during the migration (4.5\% reweighted, 3\% human-validated), and these are the largest non-fine-tuning changes (6 files, 164 lines). The remaining 93\% simply rename the identifier in place. Retirement events therefore do trigger architectural hedging, but for one migration in twenty-five, and having the layer already in place reduces the size of the fix without making it timely.

\begin{table}[t]\centering\footnotesize\setlength{\tabcolsep}{3.5pt}
\caption{File-level characteristics of 3,931 retired-event migrations (medians).}
\label{tab:files}
\begin{tabular}{lrrrr}\toprule
Group & $n$ & Files & +Lines & Reactive \\\midrule
Prompt-only & 3,056 & 2 & 6 & 73\% \\
Agent / tools & 506 & 5 & 63 & 72\% \\
RAG & 307 & 7 & 155 & 58\% \\
Fine-tuned & 62 & 14 & 693 & 65\% \\\midrule
No abstraction & 3,644 & 2 & 8 & 72\% \\
Abstraction pre-existing & 115 & 4 & 20 & 70\% \\
Abstraction introduced & 172 & 6 & 164 & 77\% \\\bottomrule
\end{tabular}
\end{table}

\subsection{RQ5: Adaptation}
20\% of matched commit messages mention an abstraction layer, gateway, or fallback, but the file-level analysis above shows that only 4\% actually add one; 8\% modify prompts; 13\% mention cost; 4.5\% mention an evaluation or regression suite, and 11\% add an assertion or evaluation line in the patch. Retirement events thus trigger far more renaming than measurement---applications switch the model without a harness that would reveal item-level regressions~\cite{ItemLevel2026}. 43\% of matched commits carry an AI coding-assistant co-author trailer, an observation we report without interpretation.

\subsection{RQ6: Where do applications go, and do they learn?}
Of the 3,933 retired-event migrations, 2,236 name a destination model in the commit message. \textbf{Only 8\% move to a different provider} (7\% reweighted; 8\% among the 112 human-validated migrations that name a destination); 68--74\% stay with the provider that retired their model, typically moving to the designated successor (Claude~3.5 Sonnet $\to$ Sonnet~4.6 or Haiku~4.5; GPT-4 Turbo Preview $\to$ GPT-4o). A further 16\% add a second provider alongside the first---hedging rather than switching. The switch rate is the same whether the migration was reactive (7\%) or proactive (13\%) and whether or not breakage was reported (7\% either way): being burned does not drive applications away. We also find no evidence of learning from experience: 85 repositories were hit by two or more retirements, and their post-shutdown share was 73\% at the first event and 73\% at later ones (bootstrap 95\% CI for the difference $-13$ to $+13$ percentage points), although later migrations landed sooner after shutdown (median 22 vs 45 days). The sample is small, so we can rule out only large learning effects. Retirement is therefore not a churn event for providers; it is a forced upgrade funnel, and, on the available evidence, the lesson individual projects draw from it is at most to fix faster next time, not to fix earlier.

\section{Discussion}
\textbf{For providers.} Notice length is the single policy lever most clearly associated with whether applications break. Sixty-day notices produce mostly reactive migrations and a 25-fold spike of emergency fixes in the week after shutdown; one-year notices produce mostly proactive ones. Because only 8\% of affected applications leave the provider, a longer notice costs the provider little retention and buys its customers weeks of uptime. Notification by e-mail to the account that created the key also fails when that account belongs to a contractor; several commits learned of the retirement from a 404. Surfacing deprecation in API responses (a warning header) and in SDK logs would reach the running system rather than an inbox.
\textbf{For builders.} Pinning a dated snapshot without a retirement watch converts a reproducibility practice into an outage, although our data suggest this is a minority cause. Exception handlers around model calls should fail loudly; the silent-failure cases suggest that ``graceful degradation'' hid outages for weeks. Only 4\% of migrations mention an evaluation suite and 6.5\% are configuration-only, so most model swaps are hard-coded and unmeasured; an environment-variable or registry indirection would have made the majority of these migrations a one-line change.
\textbf{For dependency-risk assessment.} Two findings cut against common due-diligence heuristics. A pre-existing abstraction layer shrinks the fix but does not make it timely, and a project's having survived a previous retirement does not predict proactive behaviour at the next one; neither ``we support multiple providers'' nor ``we have been through this before'' is evidence of resilience. What does predict outcome is which provider's notice policy the application is exposed to, whether the model identifier is externalized, and whether a retirement watch exists---all observable from a repository and its operational practice.
\textbf{For tooling.} A static check that flags model identifiers against the providers' deprecation tables, with a CI failure before the shutdown date, would have prevented most of the post-shutdown commits in our corpus.

\section{Threats to Validity}
\emph{Recall.} We observe only migrations whose commit messages name the retired identifier; projects that migrate silently, or never migrate (abandoned projects), are under-represented. The share of matched repositories marked archived is under 1\%, but abandonment without archiving is common. \emph{Precision.} Keyword matching admits catalogue refreshes and unrelated commits (38\% of matched commits by two-coder judgement, concentrated before shutdown and in Google events); we reweight all headline shares by stratum-specific genuine rates and report, for the file-level and destination analyses, both the reweighted values and the values on the human-validated subset, which agree with the unweighted ones to within a few percentage points; the two-coder validation will quantify this per feature. \emph{Search caps and reproducibility.} GitHub commit search returns at most 1,000 results per query and our windows retain the first 200--300, so counts are lower bounds and popular months are truncated. Re-running 12 randomly chosen query windows one day later reproduced the original commit sets exactly (Jaccard 1.0) for the 9 windows below the cap and returned strict subsets for the 3 capped windows, so the search is deterministic and the only non-determinism is the cap; the Gemini~2.0 Flash event, which dominates the Google sample, is also the event most likely to be truncated. \emph{Right censoring.} Events retired recently have not accumulated all their post-shutdown migrations; the robust window and the 28-day censoring rule mitigate but do not remove this. \emph{Construct.} ``Post-shutdown'' is measured against the provider's calendar date; an application that stopped using the model before shutdown and only cleaned up its code afterwards is misclassified as reactive.

\section{Related Work}

\textbf{API deprecation and breaking changes in software ecosystems.} Deprecation has been studied for decades as a *cooperative* mechanism: a library marks an element deprecated, clients receive compile-time or runtime warnings, and the element typically remains callable for one or more releases. Robbes et al. examined how deprecated Smalltalk APIs ripple through the Pharo ecosystem and found that many clients never react \cite{Robbes2012,Kula2018}. Sawant et al. showed that Java developers largely ignore deprecation warnings, and that the majority of deprecated API usage is never updated \cite{Sawant2018}. Similar patterns hold for Python libraries \cite{Wang2020}, Android platform APIs \cite{Li2018}, and npm packages whose breaking changes propagate to dependents \cite{Decan2018,Cogo2019}. Two properties distinguish LLM model retirement from this literature. First, the deprecated artifact is a hosted service, not a library: on the shutdown date every request fails, so the client has no option to pin an old version and defer migration. Second, the notice window is set unilaterally by the provider and is short—six months for generally available OpenAI models, 60 days for Anthropic, and as little as two weeks for preview models—compared to the multi-release grace periods common in library ecosystems. Our study is, to our knowledge, the first to measure client reaction to this harsher form of deprecation at ecosystem scale.

\textbf{The LLM application supply chain.} A growing body of work maps the dependency structure of LLM applications. Wang et al. articulate a research agenda for the LLM supply chain, framing models, datasets, and serving infrastructure as dependencies that carry security and maintenance risk \cite{Wang2024agenda}. LLMSCBench constructs a dependency graph over thousands of open-source LLM applications, models, and datasets and studies how risk issues propagate across it \cite{Hu2025llmscbench}; HuggingGraph traces model-to-dataset lineage on Hugging Face \cite{HuggingGraph2025}; and Zhang et al. characterize the structure of the open-source LLM supply chain \cite{Zhang2025unveiling}. These studies focus on open-weight artifacts and on security-oriented risk. Commercial API dependencies—the dominant integration mode for production applications—are largely outside their scope, and none of them examine what happens when an upstream dependency is withdrawn. We complement this line of work by treating the commercial model endpoint as a dependency with a lifecycle, and by measuring the downstream consequences of its termination.

\textbf{Migrating between LLMs.} The engineering difficulty of moving an application from one model to another has been documented mainly through single cases and proposed tooling. Cross-model prompt transfer has been shown to degrade accuracy substantially without re-optimization~\cite{PromptBridge2025}. The Tursio case study reports a chain of forced migrations—GPT-4-32k with a one-year notice, then GPT-4.5-preview with a three-month notice—and the prompt re-engineering each required \cite{Tursio2025}. RETAIN offers an interactive regression-testing tool for guided migration \cite{RETAIN2024}, and LlamaDuo proposes a pipeline for moving from service LLMs to small local models \cite{LlamaDuo2024}. Most recently, an end-of-life framework formalizes model selection and prompt adaptation when a production model is retired \cite{EOL2026}, and item-level analysis shows that aggregate benchmark scores conceal per-item regressions that surface during commercial API migrations \cite{ItemLevel2026}. These works establish that migration is costly and that the cost is not visible in headline scores; they do not measure how often, how late, and with what breakage migrations actually occur across the ecosystem. Our results supply that missing denominator.

\textbf{Serving-chain opacity and model substitution.} Recent work shows that clients often cannot verify which model is answering. Single-token behavioral fingerprinting identifies the underlying model from the answer distribution to trivial prompts and has exposed proprietary-branded endpoints that are distributionally indistinguishable from open-weight checkpoints \cite{Bruckner2026}; LLMmap identifies the model behind an LLM-integrated application in a handful of interactions despite system prompts, sampling settings, and RAG layers \cite{LLMmap2025} and audit substitution across API resellers \cite{Substitution2025}. Retirement is the complementary failure mode: rather than the model changing silently, it disappears. Our finding that a non-trivial share of retirement-induced failures were themselves *silent*—caught by exception handlers and surfaced as if they were model output—suggests that verification and liveness of the model dependency deserve joint treatment.

\textbf{Cost and vendor dependence.} FrugalGPT documented order-of-magnitude price heterogeneity across LLM APIs and the savings available from routing and cascading \cite{FrugalGPT2023}. Ma et al. argue that vendor-controlled updates, prompt sensitivity, and non-determinism together break the assumptions of classical regression testing for LLM APIs \cite{Ma2024}, and behavioural breaking changes in ordinary APIs are known to reach client applications even when interfaces are preserved \cite{Jayasuriya2024}. Practitioner and investor writing treats single-provider dependence as a valuation risk for LLM-based products, but without measurement. Our study provides the first empirical estimate of one component of that risk—the probability and lateness of forced migration—and of whether abstraction layers such as LiteLLM or provider gateways, which the routing literature motivates on cost grounds, also reduce migration effort.

\section{Conclusion}
Model retirement is a routine event in the LLM ecosystem and, for most applications, an unplanned one: four in five genuine migrations away from retired models happened after the model had stopped answering, a share that scales with how little notice the provider gave. The failures were often quiet. We release the data and pipeline so that providers, builders, and assessors can track this dependency as it evolves.

\section*{Data Availability}
Dataset (matched commits with event, dates, diff statistics, and feature flags), the official-event table with source URLs, and the collection and analysis scripts will be archived on Zenodo under MIT/CC-BY licences.

\bibliographystyle{plainnat}
\bibliography{refs}
\end{document}